\documentclass{moriond}

\usepackage{amsmath,amssymb} 
\usepackage{siunitx}

\def\Journal#1#2#3#4{{#1} {\bf #2}, #3 (#4)}

\def\lemaitre{\textsc{Lema\^itre}\ }
\def\dd{\mathrm{d}}

\def\be{\begin{equation}}
\def\ee{\end{equation}}
\def\bea{\begin{eqnarray}}
\def\eea{\end{eqnarray}}

\newcommand{\Photo}{}

\begin{document}
\vspace*{4cm}
\title{Accurate measurement of telescope filter bandpasses with a Collimated Beam Projector and impact on cosmological parameters}

\author{J\'er\'emy Neveu$^{1,2}$, Dylan Kuhn$^{1}$, Thierry Souverin$^{1}$, on behalf of the \lemaitre collaboration}

\address{$^{1}$Sorbonne Universit\'e, CNRS, Universit\'e de Paris, LPNHE, 75252 Paris Cedex 05, France; $^{2}$Universit\'e Paris-Saclay, CNRS, IJCLab, 91405, Orsay, France.}

	
\maketitle\abstracts{
The measurement of magnitudes with different filters in photometric surveys gives access to cosmological distances and parameters. However, for current and future large surveys like the ZTF, DES, HSC or LSST, the photometric calibration uncertainties are almost comparable to statistical uncertainties in the error budget of type Ia cosmology analysis, which limits our ability to use type Ia supernovae for precision cosmology. The knowledge of the bandpasses of the survey filters at the per-mill level can help reach the sub-percent precision for magnitudes. We show how a misknowledge of the bandpasses central wavelengths or of the presence of out-of-band leakages leads to biased cosmological measurements. Then, we present how to measure the filter throughputs at the required precision with a Collimated Beam Projector.
}


To probe the Universe dynamic and understand the nature of dark energy, cosmological photometric surveys need to compute distances from the measurement of the colors of astrophysical sources. Then, because SN Ia colours are redshifted with universe expansion, high redshift supernovae are measured through telescope redder bands while low redshift supernovae are observed in bluer bands. This case underlines that colours need to be accurately calibrated in an optical survey to have the correct distance ratios between the SN observations. Every chromatic effect from the instrumental response distort our dynamic perception of the universe expansion.

Hubble-Lema\^itre diagram uses the rest-frame B-band apparent magnitude $m_B^*(z)$, the standardized SN flux, to get cosmological parameters. This magnitude is built from observed magnitudes $m_X$ in band $X$, involving colour transformations, redshift, and astrophysics:
\begin{equation}
m_B^*(z) = \underbrace{m_X - K_{XB}(z)}_{observations} = \underbrace{\mu(z)}_{cosmology} + \underbrace{M_B + \alpha x_1 + \beta c + \Delta M_{\text{host}}}_{astrophysics}
\end{equation}
with $\mu(z)$ the distance modulus that contains the cosmological model, $M_B+\alpha x_1 + \beta c + \Delta M_{\text{host}}$ is the SN Ia absolute magnitude in band $B$. The $K$-correction $K_{XB}(z)$ represents the correction in magnitude that would have to be made if the star were observed in its reference frame at rest. Importantly, it mixes the redshift $z$ and the knowledge of transmissions:
\begin{equation}
K_{XB}(z)  = -2.5\log_{10} \left[\frac{1}{(1+z)} \frac{\int \lambda \dd \lambda F_{\mathrm{ref}}(\lambda)B(\lambda) \int \lambda \dd \lambda L_\lambda(\lambda/(1+z)) T_X(\lambda)T_{\mathrm{atm}}(\lambda) }{\int \lambda \dd \lambda F_{\mathrm{ref}}(\lambda) T_X(\lambda)T_{\mathrm{atm}}(\lambda)\int \lambda \dd \lambda L_\lambda(\lambda) B(\lambda)} \right]
\end{equation}
with the spectral density $F_\lambda$ of the supernova, to be constructed by a spectrophotometric model fitted to the measured spectral sequences; the reference spectral density $F_{\mathrm{ref}}(\lambda)$, to be established by measurements or stellar atmosphere modelling; the transmission of the telescope filters $T_X(\lambda)$; and the atmospheric transmission of the observation site $T_{\mathrm{atm}}(\lambda)$. Every factor must be well-known in order to measure $w$ at the sub-percent level.


\lemaitre project (standing for Latest Extended Mapping of Acceleration with an Independent Trove of
  Redshifted Explosions) is the combination of 3 new and independent surveys (ZTF-II, SNLS 5yr and HSC) of $\approx 3500$ spectroscopically confirmed SNe~Ia, anchored on an instrumental photometric calibration from \textsc{StarDICE} \cite{stardice}. We investigated the impact on $w$, the dark energy equation of state parameter, fitted on a noise-free simulation of the \lemaitre survey from inaccuracies on the filter bandpass estimates. We did not re-trained the SALT2 in the following toy model: effects of filter miscalibration may be worse with training!
  
\begin{figure}[h]
\centering
\includegraphics[width=0.48\linewidth]{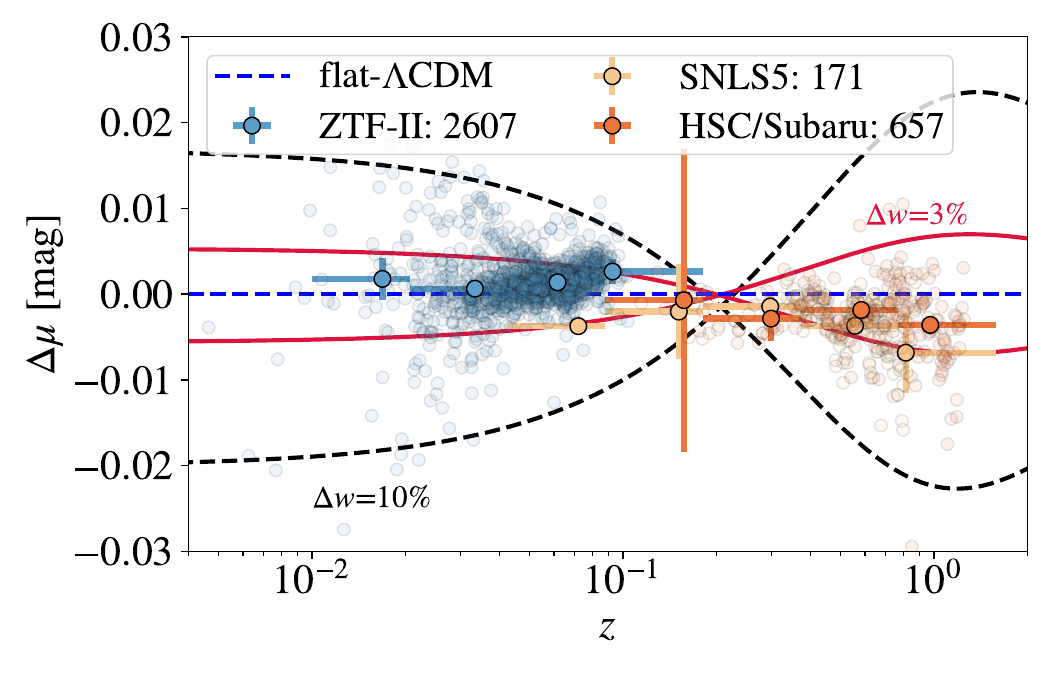}\hfill
\includegraphics[width=0.48\linewidth]{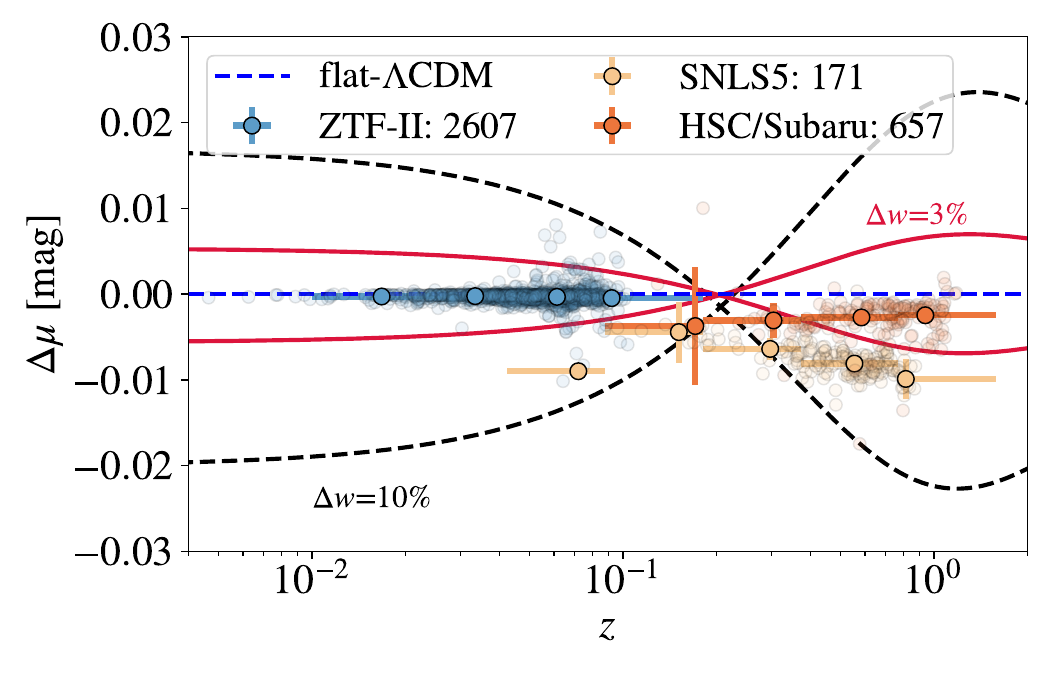}
\vspace{-1.2em}
\caption{Left: Hubble diagram residuals $\Delta \mu$ due to ZTF filters shifted by $\SI{-5}{\angstrom}$, others by $\SI{5}{\angstrom}$. Right: Hubble diagram residuals $\Delta \mu$ due to additional blue leakages in $i$ and $z$ filters of $10^{-3}$ peak amplitude. Red (resp. dashed black) curves corresponds to $\mu$ variations if $w$ is varied from 3\% (resp. 10\%).}\label{fig:ztfShift}
\end{figure}

\textbf{First scenario:} the ZTF filters were shifted by $\SI{-5}{\angstrom}$ and the others by $+\SI{5}{\angstrom}$. We observed a distortion of the Hubble diagram, which can lead to a deviation on $w$ of about 2\% (Figure \ref{fig:ztfShift} left). The future goal to get $\Delta w \lesssim 1\%$ requires to measure the filter positions at the $\SI{1}{\angstrom}$ level.

\begin{figure}[h]
\centering
\includegraphics[width=0.42\linewidth]{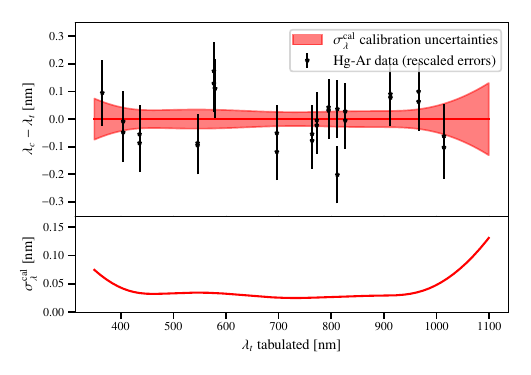}\hfill
\includegraphics[width=0.58\linewidth]{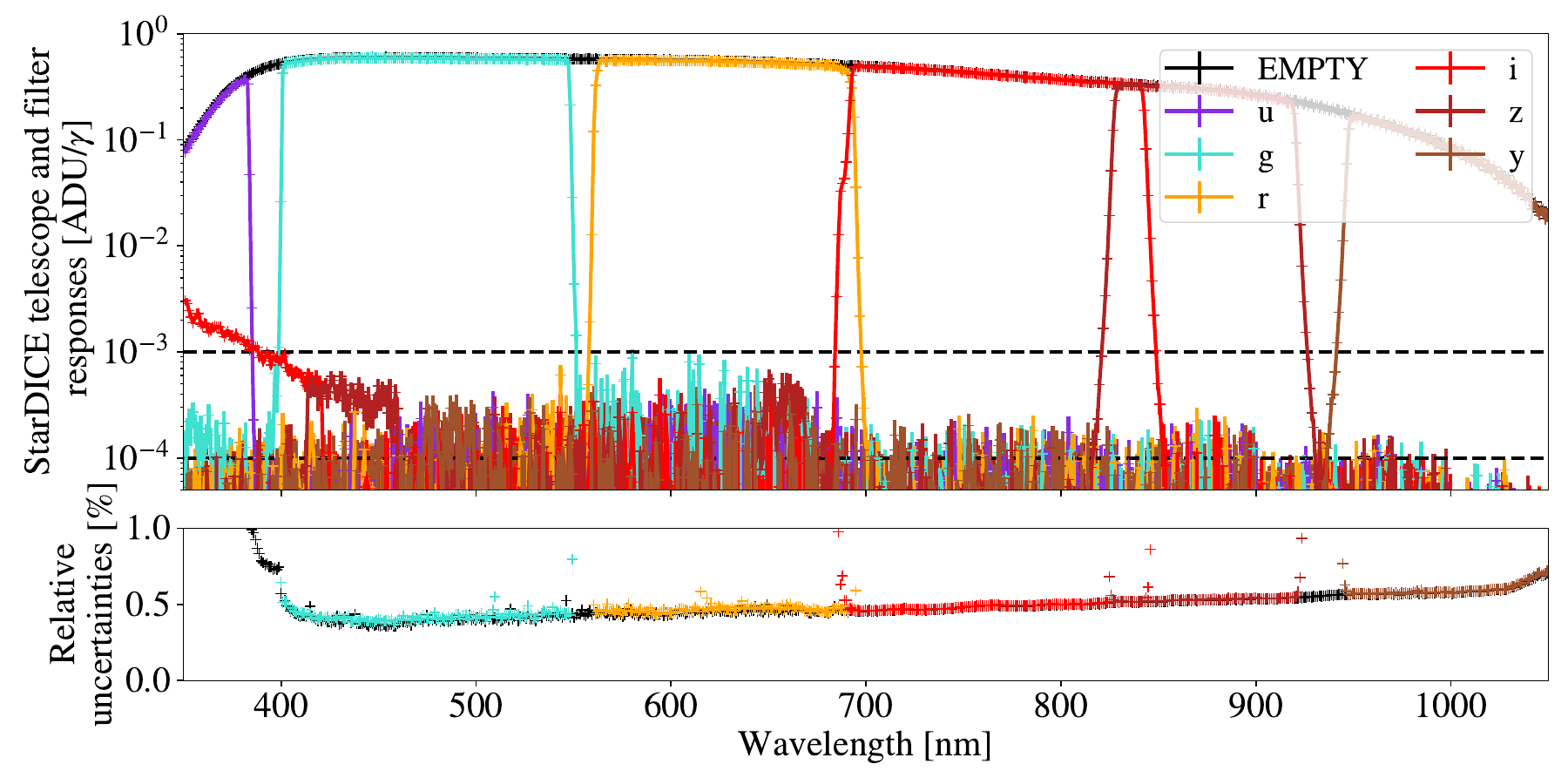}
\vspace{-1.8em}
\caption{Left: spectrograph wavelength calibration using a Hg-Ar lamp and uncertainties. Right: measurement of the \textsc{StarDICE} $ugrizy$ filters: blue leaks at the $10^{-4}$ level are clearly detected in the \textsc{StarDICE} red filters.}\label{fig:cbp}
\end{figure}

\textbf{Second scenario:} we added out-of-band leakages of peak amplitude $10^{-3}$ at $\approx\SI{450}{\nm}$ to the original $iz$ filters of all surveys (like in Figure~\ref{fig:cbp}). We observed that $w$ can be bias by $\approx 2\%$ (Figure~\ref{fig:ztfShift} right), then filter leakages must be checked with high precision on the full visible range.

The Collimated Beam Projector (CBP) is a parallel beam of monochromatic light monitored in flux and wavelength to measure $T_X(\lambda)$ \cite{CBP}. We built a CBP with a tunable laser source and a reversed telescope, and we tested its performance on the measurement of the \textsc{StarDICE} telescope filters. 
In the lab, we succeeded in calibrating the CBP throughput at 0.5\%, systematics included (mostly due to scattered light), and in wavelength at $\approx \SI{0.2}{\angstrom}$. Statistical noise is low-enough to detect $5\times 10^{-4}$ blue leakages in the red filters. 
The CBP is thus a new device that is accurate enough to calibrate next-generation cosmological photometric surveys.

\end{document}